\documentclass[twocolumn,showpacs,preprintnumbers,amsmath,amssymb,prb,superscriptaddress]{revtex4-1}
\usepackage{bbm}
\usepackage{mathrsfs}
\usepackage{graphicx}
\usepackage{dcolumn}
\usepackage{bm}
\usepackage{amsmath}
\usepackage{amsfonts}
\usepackage{color}
\usepackage[colorlinks=true,citecolor=blue,anchorcolor=blue]{hyperref}
\usepackage{floatrow}

\begin{document}

\title{Gapless Higgs Mode in the Fulde-Ferrell-Larkin-Ovchinnikov State of a Superconductor}

\author{Zhao Huang}
\affiliation{Theoretical Division, Los Alamos National Laboratory, Los Alamos, New Mexico 87545, USA}
\affiliation{Texas Center for Superconductivity, University of Houston, Houston, Texas 77204, USA}

\author{C. S. Ting}
\affiliation{Texas Center for Superconductivity, University of Houston, Houston, Texas 77204, USA}

\author{Jian-Xin Zhu}
\email{jxzhu@lanl.gov}
\affiliation{Theoretical Division, Los Alamos National Laboratory, Los Alamos, New Mexico 87545, USA}

\author {Shi-Zeng Lin}
\email{szl@lanl.gov}
\affiliation{Theoretical Division, Los Alamos National Laboratory, Los Alamos, New Mexico 87545, USA}

\begin{abstract}
The Higgs mode associated with amplitude fluctuations of the superconducting gap in uniform superconductors usually is heavy, which makes its excitation and detection difficult. We report on the existence of a gapless Higgs mode in the  Fulde-Ferrell-Larkin-Ovchinnikov {states}. {This} feature is originated from the Goldstone mode associated with the translation symmetry breaking. The existence of the gapless Higgs mode is demonstrated by using both a phenomenological model and microscopic Bardeen–Cooper–Schrieffer (BCS) theory. The gapless Higgs mode can avoid the decay into other low energy excitations, which renders it stable and detectable.
\end{abstract}
\date{\today}
\maketitle

\date{\today}

%\pacs{
%        73.22.-f  % Electronic structure of nanoscale materials and related systems
%        02.20.-a  % Group theory
%        73.43.-f  % Quantum Hall effects
%      }

{\it Introduction ---} The idea of spontaneous symmetry breaking, that a state does not need to have the same symmetries as the model Hamiltonian that describes the system under consideration, is one of the cornerstones of modern physics. Perhaps the most important example is the breaking of the gauge symmetry related to the weak and electromagnetic interaction $\mathrm{SU(2)}\times\mathrm{U(1)}$, which creates the Higgs condensate as the origin of particle masses in the Standard Model. The corresponding Higgs boson, which is generated by the quantum excitation of the condensate, was discovered experimentally in 2012 \cite{PLB1, PLB2}. The Higgs boson as an elementary particle has a large mass thus requiring a huge particle collider, such as CERN at Europe, to enable its discovery. 

An elementary excitation, analogous to Higgs boson, can also appear in condensed matter systems \cite{Littlewood81, Littlewood82,yusupov_coherent_2010,PhysRevB.26.4883,PhysRevLett.115.157002,sherman_higgs_2015,doi:10.1146/annurev-conmatphys-031119-050813,PhysRevResearch.2.013034,PhysRevB.46.8934,PhysRevLett.119.067201,jain_higgs_2017,volovik2014higgs, Tsuchiya18}. One example is the amplitude mode in superconductors, which is associated with the amplitude fluctuations in the superconducting condensate from breaking of the U(1) gauge symmetry \cite{Littlewood82}. This mode is widely referred as the Higgs mode in literatures \cite{doi:10.1146/annurev-conmatphys-031119-050813}. Recently, the Higgs mode has been observed in conventional $s$-wave superconductors ($\mathrm{Nb_x Ti_{1-x} N},\ \mathrm{NbN}$) by ultrafast terahertz (THz) pump-THz probe spectroscopy \cite{Matsunaga13, Matsunaga14}, which revives the interest in the Higgs dynamics of superconducting order parameter. Akin to its cousin in particle physics, the Higgs boson in superconductors is very massive, which renders it short lived by decaying into quasiparticle continuum. Lots of efforts have been made to reduce the energy of the Higgs mode, and in certain circumstance, the energy of the Higgs mode can be made smaller than the continuum of the excitations, which results in a stable Higgs mode~\cite{hong_higgs_2017,PhysRevLett.122.127201,PhysRevB.102.125102}.

In this work, we demonstrate the existence of a gapless Higgs mode in the Fulde-Ferrell-Larkin-Ovchinnikov (FFLO) state of a superconductor.  This is based on the observation that the FFLO state breaks spatial translational invariance, thus admitting a gapless ``phonon" mode, which is just the Higgs mode.  The damping of this low energy Higgs mode is suppressed and the Higgs mode has a long lifetime. The FFLO state was predicted in Pauli limited superconductors due to the imbalanced up and down spin species under a magnetic field \cite{FF64,LL64}. The superconducting gap function $\Delta(r)=\langle\psi_\uparrow \psi_\downarrow\rangle$ has a nonzero center-of-mass momentum and oscillates in space. There are two competing FFLO states:  the state with only phase modulation is called the FF state \cite{FF64}; the state with only amplitude modulation is  called the LO state \cite{LL64}. It has been found that the LO state has a lower free energy. Possible FFLO states have been reported experimentally in layered organic superconductors, heavy Fermion superconductors and $\mathrm{FeSe}$ \cite{RMP04, Matsuda07,PhysRevLett.91.187004,FFLONMR2006,PhysRevLett.107.087002,PhysRevLett.109.027003,mayaffre_evidence_2014,PhysRevLett.116.067003,PhysRevLett.121.157004,PhysRevLett.119.217002,PhysRevB.85.174530,PhysRevB.97.144505,PhysRevLett.118.267001,PhysRevLett.124.217001}.

{\it Effective low-energy theory ---} In the FFLO state, the spatial translational symmetry is spontaneously broken, which gives rise to a new type of Goldstone mode. The energy of the system is invariant $E(\Delta(\mathbf{r}))=E(\Delta(\mathbf{r}+\mathbf{\delta_r}))$ under a spatial translation $\mathbf{\delta_r}$, where $\Delta(\mathbf{r})$ is a spatial dependent order parameter. The eigenstate of the Goldstone mode is $\partial_r\Delta$. In  the FF state, the Goldstone mode is the phase mode, while in the LO state, the Goldstone mode is the amplitude or Higgs mode associated with the amplitude fluctuations. In this paper, we focus on the Higgs mode in the LO state. The relation between the Goldstone mode of the FFLO state and the Higgs mode can be understood using a Ginzburg-Landau action. To the second order in superconducting order parameter $\Psi$, the low-energy effective Lagrangian can be written as \cite{buzdin_generalized_1997}
\begin{eqnarray}\label{eq1}
\!\!\!\!\!\mathcal{L}\!=\!i K_1\Psi^*\partial_t \Psi\!+\!K_2|\partial_t\Psi|^2\!\!-\!\![\alpha |\Psi|^2 \!+\! \gamma |\nabla\Psi|^2\!+\!\eta |\nabla^2\Psi|^2].
\end{eqnarray}
Here we have neglected the coupling of the Cooper pairs to the electromagnetic field for convenience of discussion. To stabilize the FFLO state, we choose $\gamma<0$ and $\eta>0$. In $\mathcal{L}$, both first and second order time derivative terms are allowed by time reversal symmetry and gauge invariant. By performing the  particle-hole transformation, $\Psi \rightarrow \Psi^*$, one can see the $K_1$ ($K_2$) term breaks (preserves) particle-hole symmetry. It is worth noticing that this particle-hole symmetry is a symmetry of the energy bands in the normal state. The particle-hole symmetry has important consequence on the collective excitation in superconductors~\cite{Pekker15}. Rigorously speaking, a pure Higgs mode only exists in superconductors with particle-hole symmetric band structure in the normal state. When the particle-hole symmetry is violated, the phase and Higgs mode start to hybridize. 

Close to $T<T_c$,  $\Psi = \Psi_0 \cos(\mathbf{Q}\cdot\mathbf{r}) e^{i\phi}$, which breaks the spatial translational invariance. The eigenstate of the corresponding Goldstone mode is $\mathbf{Q}\cdot\mathbf{\delta_r}\sin (\mathbf{Q}\cdot\mathbf{r})$ with $\mathbf{\delta_r}$ the small spacial shift, which is just the Higgs mode of the superconducting order parameter. Therefore the breaking of the spatial translational invariance in the FFLO state guarantees the existence of the gapless Higgs mode. In the presence of particle-hole symmetry ($K_1=0$), the dispersion of the Higgs mode is {$\Omega_H=v_H q$} according to Eq.~\eqref{eq1} since the inversion symmetry of the Lagrangian restricts the lowest order of $q$ as $q^2$, same as the order of $\Omega_H$ from $|\partial_t\Psi|^2$. This mode can also be regarded as the ``phonon" mode of the crystal of the superconducting order in the FFLO state. The presence of impurities can gap the Higgs mode by a pinning of the FFLO order parameter. We will consider clean systems in the following discussions. It is worth mentioning that a Higgs mode being gapless or gapful has no direction connection with the superconducting order parameter in the momentum space being nodal or nodeless. For example, spatially uniform $d$-wave superconductors with nodal lines have a gapful Higgs mode \cite{Varma13,Peronaci15,Schwarz20}.

In the FFLO state, there are nodal regions with $|\Psi(r)|=0$, where the local quasiparticles become gapless. The Higgs mode can decay into these quasiparticles, which renders the lifetime of the Higgs mode being finite. The decay process conserves energy, and as the Higgs mode energy approaches $\Omega_H\rightarrow 0$, the phase space of decay is reduced significantly. It is expected that the Higgs mode becomes long lived in that limit.

\begin{figure}[t]
\begin{center}
\includegraphics[clip = true, width =\columnwidth]{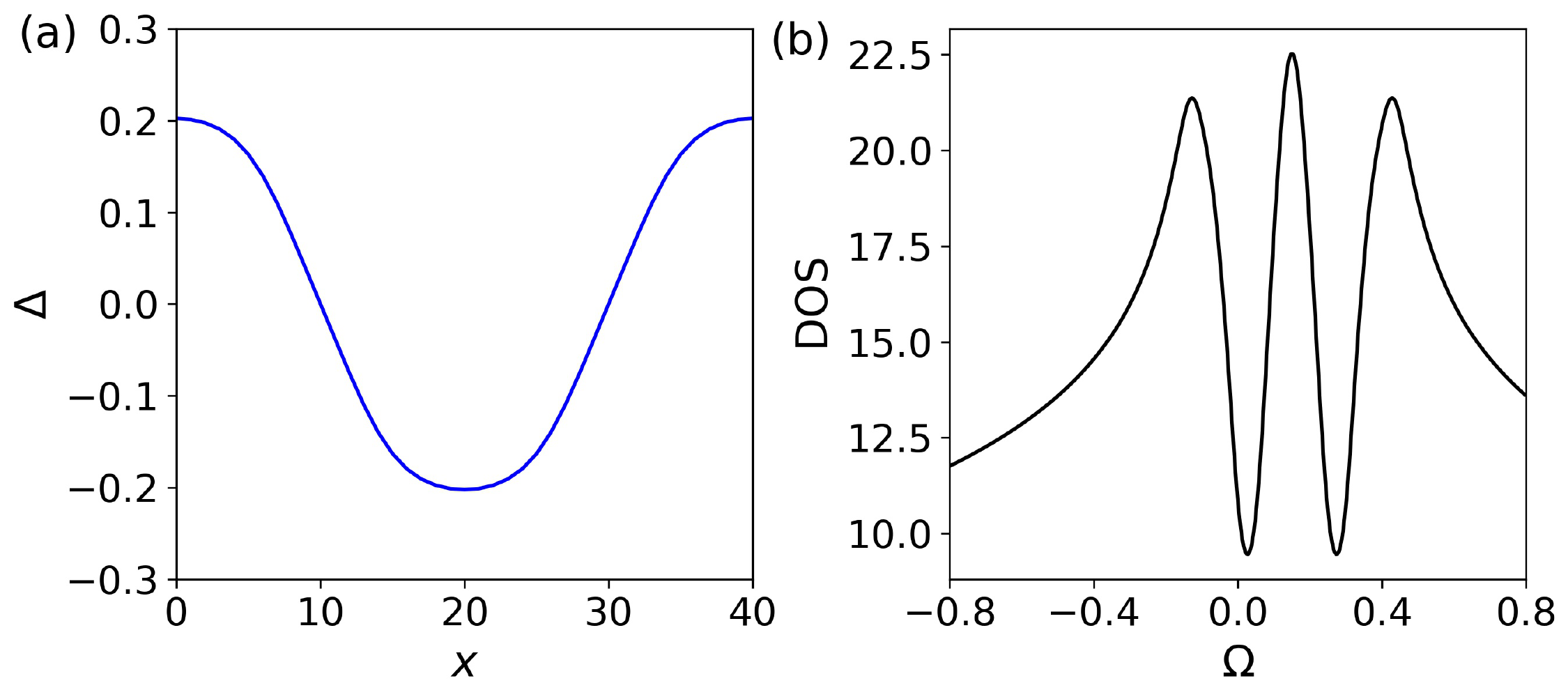}
\caption{(a) Spatial dependence of superconducting order parameter in the FFLO state, and (b) the corresponding density of state. The parameters are $V=1.6$, $E_z=0.15$, $t=1$.}
\label{fig1}
\end{center}
\end{figure}

{\it Field theoretic approach ---} To gain further insight and to demonstrate explicitly the damping of the Higgs mode, we study the Higgs mode using the microscopic BCS theory. We consider a superconducting film made of an $s$-wave superconductor under a parallel magnetic field. In this case, the orbital coupling is absent and the upper critical field is limited by the Pauli pairing breaking effect.
The Hamiltonian is given by
\begin{eqnarray}\label{eq3}
H=\int dr^2 \psi_{\mathbf{r}\sigma}^\dagger \biggl{[}\biggl{(}\frac{-\nabla^2}{2m}-\mu\biggr{)}\delta_{\sigma\sigma'} +E_z\sigma_{z,\sigma\sigma'}\biggr{]}\psi_{\mathbf{r}\sigma'}  \nonumber \\
-V \int dr^2 \psi_{\mathbf{r}\uparrow}^{\dagger}\psi_{\mathbf{r}\downarrow}^{\dagger}\psi_{\mathbf{r}\downarrow}\psi_{\mathbf{r}\uparrow},
\end{eqnarray}
with $\mu$ the chemical potential, $E_z$ the Zeeman term and $V$ the pairing interaction strength. We have taken the unit $\hbar=1$. Using the standard Hubbard-Stratonovich transformation, we obtain the action with superconducting gap function $\Delta$ in the imaginary time domain \cite{Atland10,PhysRevLett.108.177005,supp}
\begin{equation}
S=\int d\tau\int d^{2}r\left(\frac{1}{V}\left|\Delta\right|^{2}\right)-\mathrm{tr}\mathrm{ln}G_{0}^{-1},
\end{equation}
with the Gorkov Green's function
\begin{equation}
G_{0}^{-1}=-\left[\begin{array}{cc}
\partial_{\tau}+h+E_{z} & -\Delta\\
-\Delta & \partial_{\tau}-h+E_{z}
\end{array}\right] \;,
\end{equation}
with $h=-\nabla^2/2m-\mu$.
We choose a gauge by setting $\Delta$ to be real by noting that the LO is the ground state in the FFLO state. In the weak coupling limit with $\Delta \ll \omega_c\ll E_F$, where $\omega_c$ is the cutoff frequency and $E_F$ is the Fermi energy, the particle-hole symmetry in the normal state is approximately preserved in the energy window $-\omega_c\le E \le \omega_c$ even in the presence of Zeeman field. The hybridization between the phase mode and Higgs mode is negligible \cite{Pekker15, Shimano20} and we will focus only on the Higgs mode in the following calculations.

We then consider the amplitude fluctuation around the saddle point solution $\Delta_0$, with $\Delta=\Delta_0+s\left(\tau,\mathbf{r}\right)$
\begin{equation}
S=\int d\tau\int d^{2}r\left(\frac{1}{V}\left|\Delta+s\right|^{2}\right)-\mathrm{tr}\mathrm{ln}\left(G_{0}^{-1}+\Sigma\right),
\end{equation}
with $\Sigma=s\sigma_{x}$. By expanding the action to the second order of $s$, we obtain the action for the fluctuation \cite{supp}  
\begin{equation}
S_{2}=\int d\tau\int d^{2}r\left(\frac{1}{V}s^{2}\right)+\frac{1}{2}\mathrm{tr}\left(G_{0}\Sigma G_{0}\Sigma\right).\label{secondorderaction}
\end{equation}
For a uniform BCS state, we can represent $s$ in the frequency-momentum representation,
\begin{equation}
\int d\tau\int d^{2}r\left(\frac{1}{V}s^{2}\right)=\frac{1}{V}\sum_{l,\mathbf{q}}s\left(-\Omega_{l},-\mathbf{q}\right)s\left(\Omega_{l},\mathbf{q}\right),
\end{equation}
and
\begin{eqnarray}
&&\mathrm{tr}\left(G_{0}\Sigma G_{0}\Sigma\right) \nonumber \\
&& =\frac{T}{L^{2}}\sum_{m,l,\mathbf{k},\mathbf{q}}\mathrm{tr}\left(\left\langle \omega_{m},\mathbf{k}\right|G_{0}\left|\omega_{m},\mathbf{k}\right\rangle \left\langle \omega_{m},\mathbf{k}\right|\Sigma\left|\omega_{m}+\Omega_{l},\mathbf{k}+\mathbf{q}\right\rangle \right.\nonumber \\
&& \left.\left\langle \omega_{m}+\Omega_{l},\mathbf{k}+\mathbf{q}\right|G_{0}\left|\omega_{m}+\Omega_{l},\mathbf{k}+\mathbf{q}\right\rangle \left\langle \omega_{m}+\Omega_{l},\mathbf{k}+\mathbf{q}\right|\Sigma\left|\omega_{m},\mathbf{k}\right\rangle \right),\nonumber
\end{eqnarray}
where $\omega_{m}=\left(2m+1\right)\pi T$ and $\Omega_{l}=2l\pi T$ are the fermionic and bosonic Matsubara frequency, and $L$ is the linear size of the system. To summarize, we rewrite $S_{2}$ as
\begin{equation}
S_{2}=\sum_{l,\mathbf{q}}s\left(-\Omega_{l},\mathbf{-q}\right)M_{q}s\left(\Omega_{l},\mathbf{q}\right),
\end{equation}
where the coupling matrix $M_{q}$ is given by
\begin{equation}
M_{q}\!\!=\!\!\left[\frac{1}{V}\!+\!\frac{1}{2}\frac{T}{L^{2}}\sum_{m,\mathbf{k}}\mathrm{tr}\!\left[G_{0}\left(\omega_{m},\mathbf{k}\right)\!\sigma_{x}\!G_{0}\left(\omega_{m}\!+\!\Omega_{l},\mathbf{k}\!+\!\mathbf{q}\right)\!\sigma_{x}\!\right]\right].\label{couplingmatrix}
\end{equation}

For the uniform superconducting state, $M_q$ can be obtained analytically because $G_0$ is a two by two matrix. At the resonance condition $M_q=0$, we obtain the well known dispersion relation for the Higgs mode as $\Omega_H^2=4\Delta^2+\frac{1}{3}v_F^2 |\mathbf{q}|^2$~\cite{Littlewood81}. The Higgs mode has a gap of $2\Delta$. The phase fluctuations are gapless due to the U(1) symmetry breaking. The phase couples to the external gauge field and becomes a plasma mode with a gap of $2\Delta$ due to the Anderson-Higgs mechanism. The decay of the Higgs mode around energy $2\Delta$ to quasiparticle continuum and plasma mode thus is suppressed. However, the Higgs mode can still decay into other low lying bosonic modes such as the phonon mode, which makes the Higgs mode short lived. Indeed, the lifetime of the Higgs mode is of the order of picosecond in experiments \cite{doi:10.1146/annurev-conmatphys-031119-050813}.

{\it Higgs mode in FFLO state ---} In the FFLO state, we calculate $\Delta_0(r)$ numerically by considering a tight-binding Hamiltonian with nearest neighbor hopping on a square lattice \cite{Ting06}
\begin{gather}
\label{AA_Hamiltonian:NI}
H_{BCS}\!=\!-t\sum_{ij,\sigma}c_{i\sigma}^{\dagger}c_{j\sigma}\!+\!E_z\sum_{i,\sigma}\sigma_z c_{i\sigma}^{\dagger}c_{i\sigma}\!-\!V\sum _i c_{i\uparrow}^{\dagger} c_{i\downarrow}^{\dagger} c_{i\downarrow} c_{i\uparrow}.
\end{gather}
We have set the chemical potential to be zero to ensure particle-hole symmetry when $E_z=0$. Standard Bogoliubov-de Gennes (BdG) method is used to solve this model \cite{Zhu_2016} with the mean-field local pairing amplitude $\Delta_i = V\langle c_{i\uparrow} c_{i\downarrow}\rangle$. We then diagonalize the mean-field Hamiltonian by the Bogoliubov transformation,
$c_{i \sigma }=\sum _{n}^{\prime} \left(u_{i \sigma }^n \gamma_{n} -\sigma v_{i \sigma }^{n*}\gamma_{n}^{\dagger }\right),
c_{i \sigma }^{\dagger }=\sum_{n}^{\prime}\left(u_{i \sigma }^{n*}\gamma_n^{\dagger }-\sigma v_{i \sigma }^n \gamma_{n} \right)$, where $\gamma^{\dagger}_n$ and $\gamma_n$ are the creation and annihilation operators for Bogoliubov quasiparticle at state $n$ and the prime sign means the sum is over all positive quasiparticle state $E_n>0$. The $u$ and $v$ coefficients are obtained from the BdG equations,
\begin{equation}
\sum_{j}
\begin{pmatrix}
(-t+E_z)\delta_{\langle ij \rangle} & \Delta_i\\
\Delta_i^{*} & (t+E_z)\delta_{\langle ij \rangle}
\end{pmatrix}
\begin{pmatrix}
u_{j\uparrow}\\v_{j\downarrow}
\end{pmatrix}
=E_{n}
\begin{pmatrix}
u_{i\uparrow}\\v_{i\downarrow}
\end{pmatrix},
\end{equation}
where  $\Delta_i=\frac{V}{2}\sum_{n}u_{i\uparrow}^{n}v_{i\downarrow}^{n*}\tanh{\left(E_{n}/2k_{B}T\right)}$.

\begin{figure}[t]
\begin{center}
\includegraphics[clip = true, width =\columnwidth]{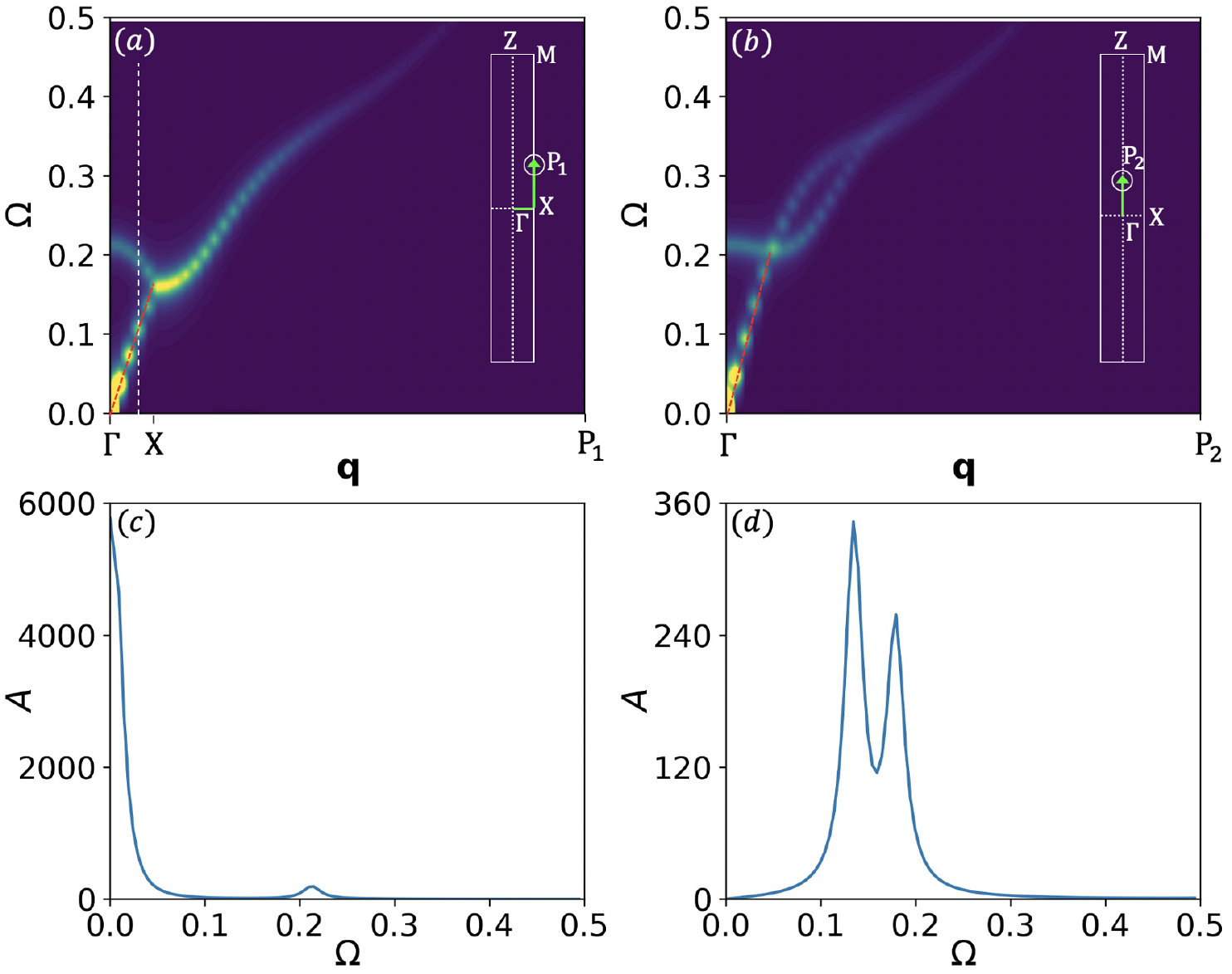}
\caption{{Contour figure of spectral density $A\left(\Omega,\mathbf{q}\right)$ of Higgs modes in the FFLO state in Fig. \ref{fig1} along the path (a) $\Gamma - {\rm X} - {\rm P}_1 $ and (b) $\Gamma - {\rm P}_2  $, where the peaks (bright yellow and green) form the spectra. (c) and (d) show the $A\left(\Omega\right)$ at $(q_x,q_z)=(0,\ 0)$ and $(0.4\pi/40a,\ 0)$ (white dash line). The damping parameter $\eta=0.01|\Delta_M|$}. The intrinsic spectra linewidth is smaller than $\eta$. The discretized bright spot (peaks) in panel (a) and (b) are due to the mesh discretization in our numerical calculations.}
\label{fig2}
\end{center}
\end{figure}

The FFLO state is calculated self-consistently by solving the BdG equations iteratively on supercells with the size $N_xa \times a$. By considering $l_x$ and $L_z$ supercells in the two directions, the total size of the thin film is $L_xa\times L_za$ with $L_x=N_xl_x$. Therefore, in the folded Brillouin zone, there are $l_x$ meshes in the $k_x$ axis and $L_z$ meshes along the $k_z$ axis. By taking $N_x=40, l_x=10$ and $L_z=400$, we find the LO state as the energy minimum for the parameters $ V=1.6t, E_z=0.15t, k_BT=0.0001$. The LO state has the maximum gap $|\Delta_M|\approx 0.22t$ at $x=0a, 20a$ as shown in Fig. \ref{fig1}(a). The total density of states (DOS) of electrons is given in Fig. \ref{fig1}(b). We can see the gap structure shifted to about $[-0.1, 0.4]$ due to the Zeeman field. The peak of DOS at $\Omega=0.15t$ is an in-gap Andreev bound state because the spacial sign change of gap function effectively plays the role of a $\pi$-junction around the node of $\Delta(r)$. The energy of Andreev bound state is shifted to $\Omega=0.15t$ due to the Zeeman coupling. This further suppresses the damping of the Higgs mode caused by decaying into Bogoliubov quasiparticles as will be shown below.

We then study the Higgs mode in the FFLO state. In principle, the Higgs mode becomes gapped when $\Delta_0(\mathbf{r})$ is commensurate with the tight-binding lattice and also in a finite system \cite{PhysRevB.99.054509}. The gap induced by pinning of the FFLO due to the tight-binding lattice is negligible because the period of the FFLO modulation is much larger than the tight-binding lattice parameter. The action for the amplitude fluctuations is \cite{supp}
\begin{equation}
S_{2}\left(\Omega,\mathbf{q}\right)=\Gamma^{\dagger}\left(-\Omega,-\mathbf{q}\right)M_{\Omega,\mathbf{q}}\Gamma\left(\Omega,\mathbf{q}\right),
\end{equation}
where $\Gamma\left(\Omega,\mathbf{q}\right)=\left[s\left(\Omega,\mathbf{q},x_{1}\right), \cdots,s\left(\Omega,\mathbf{q},x_{N_{x}}\right)\right]^{\mathrm{\dagger}}$ and
\begin{eqnarray}
\!\!\!\!\!\!\!M_{\Omega,\mathbf{q},ij}\!=\!\frac{1}{V}\!+\!\frac{1}{2L_{z}l_{x}}\sum_{\mathbf{k},n,d}\frac{D\left[f\left(E_{\mathbf{k}}^{n}\right)-f\left(E_{\mathbf{k+q}}^{d}\right)\right]}{E_{\mathbf{k}}^{n}+\Omega+i\eta-E_{\mathbf{k+q}}^{d}},
\end{eqnarray}
with $f(E)$ the Fermi-Dirac distribution function and $\eta$ being a damping parameter which is introduced for convenience of numerical evaluation. Here
\begin{eqnarray}
D &&=u_{\mathbf{k}}^{n}\left(x_{i}\right)v_{-\mathbf{k}}^{n*}\left(x_{j}\right)u_{\mathbf{k}+\mathbf{q}}^{d}\left(x_{j}\right)v_{-\left(\mathbf{k}+\mathbf{q}\right)}^{d*}\left(x_{i}\right) \nonumber \\
&&+u_{\mathbf{k}}^{n}\left(x_{i}\right)u_{\mathbf{k}}^{n*}\left(x_{j}\right)v_{-\left(\mathbf{k}+\mathbf{q}\right)}^{d}\left(x_{j}\right)v_{-\left(\mathbf{k}+\mathbf{q}\right)}^{d*}\left(x_{i}\right) \nonumber \\
 && +v_{-\mathbf{k}}^{n}\left(x_{i}\right)v_{-\mathbf{k}}^{n*}\left(x_{j}\right)u_{\mathbf{k}+\mathbf{q}}^{d}\left(x_{j}\right)u_{\mathbf{k}+\mathbf{q}}^{d*}\left(x_{i}\right) \nonumber \\
 &&+v_{-\mathbf{k}}^{n}\left(x_{i}\right)u_{\mathbf{k}}^{n*}\left(x_{j}\right)v_{-\left(\mathbf{k}+\mathbf{q}\right)}^{d}\left(x_{j}\right)u_{\mathbf{k}+\mathbf{q}}^{d*}\left(x_{i}\right),
\end{eqnarray}
where $n,d$ are the index of eigen-energy indices. The Green function of the Higgs modes is given by $G_{\Omega,\mathbf{q}}=-M_{\Omega,\mathbf{q}}^{-1}$ and the spectral density is given by  $A\left(\Omega,\mathbf{q}\right)=-\mathrm{Im}\left(G_{\Omega,\mathbf{q}}\right)/\pi$.

Figure \ref{fig2}(a) and (b) show the low-energy and long-wavelength spectra of Higgs mode (bright yellow and green) along the path $\Gamma - {\rm X}- {\rm P_1}$ and $\Gamma - {\rm P_2}$ where ${\rm P_1}$ and ${\rm P_2}$ are on the path ${\rm X}- {\rm M}$ and $\Gamma - {\rm Z}$. Because of the high anisotropic supercell with $40a\times a$, we have $\Gamma - {\rm X}$ much shorter than ${\rm X} - {\rm M}$ and $\Gamma - {\rm Z}$. The spectra are formed by the peaks of spectral function at ${\bf q}$ cuts, as shown in Fig. \ref{fig2}(c) and (d). The peak locations for small $|{\bf q}|$ values connect into a linear line starting from the $\Gamma$ point, which indicates a gapless Higgs mode, consistent with the analysis in Eqs. \eqref{eq1}. Far from the $\Gamma$ point, the peaks become more obscure with increasing $|{\bf q}|$ and finally disappear for a large $|{\bf q}|$, indicating a strong damping of the Higgs mode by coupling to the quasiparticle continuum. This means that the Higgs mode is only well defined in the long-wavelength limit.

\begin{figure}[t]
\begin{center}
\includegraphics[clip = true, width =\columnwidth]{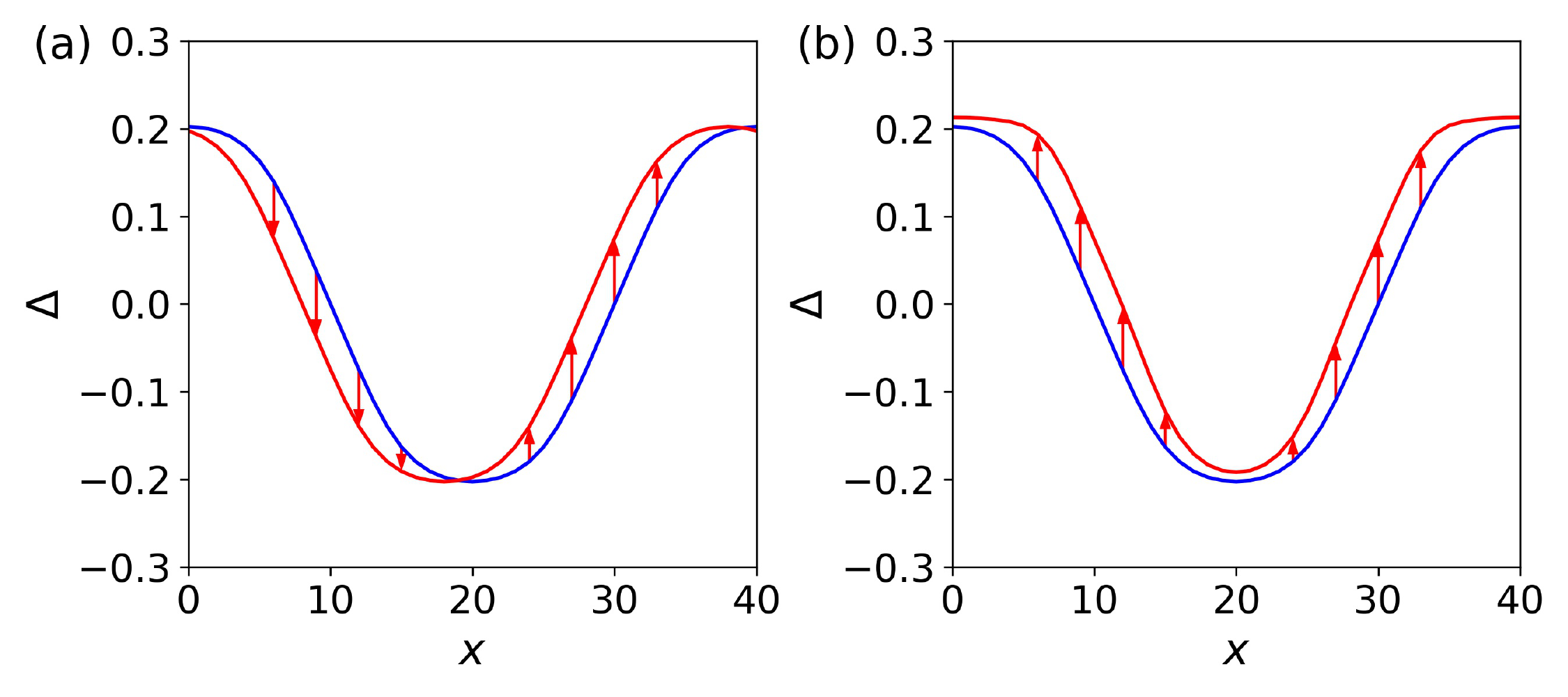}
\caption{Amplitude fluctuations of $\Delta(x)$ corresponding to the Higgs mode at (a) $\Omega=0$ and (b) $0.21$ at the $\Gamma$ point. The blue curves show the equilibrium state and the red curves show the excited states with fluctuations. The gap functions are uniform along the $z$ direction.}
\label{fig3}
\end{center}
\end{figure}

At $\Gamma$ point, except for a peak at $\Omega=0$, there is another visible peak at $\Omega\approx 0.21t$, which indicates another amplitude mode. To understand these two modes, we calculate the eigenstates of $M_{\Omega,\mathbf{q}}$ at ${\bf q}=0$, which are presented in the form of arrows in Fig. \ref{fig3}. Fig. \ref{fig3}(a) shows the gapless mode which corresponds to the spatial translation of the FFLO order parameter. Fig. \ref{fig3}(b) shows the eigenstate for the gapped mode at $\Omega\approx 0.21t$, which corresponds to the breathing of the FFLO order parameter.

The existence of the gapless Higgs mode has consequence on the stability of the FFLO state. In 3D, the FFLO state has a true long range order. However in 2D, the FFLO state becomes quasi-long-range order due to the gapless mode according to the Mermin-Wagner theorem. To induce the FFLO state, the particle-hole symmetry is not exact in the normal state because of the Zeeman field induced spin band splitting. Therefore, there is coupling between the Higgs mode and the phase mode. In charged superconductors, the phase mode becomes plasma mode with the gap proportional to the local superconducting order parameter. The decay of the Higgs mode into the phase mode is thus suppressed. In charge neutral superfluids, the phase mode is gapless. In the uniform state, the gapped Higgs mode decays quickly into the phase mode, and its visibility depends on the dimensionality of the system \cite{PodolskyArovasPRB}. Nevertheless, in the FFLO state, the gapless Higgs mode remains stable even in presence of the gapless phase model.  So far we have focused on the collective excitation associated with translation of the FFLO. Spatially localized topological excitation in the translation of the FFLO state is allowed, which corresponds to the dislocation of the FFLO order and can be regarded as localized topological Higgs mode. 

It is recognized that the FFLO state rather belongs to a broader class of order called pair density wave~\cite{doi:10.1146/annurev-conmatphys-031119-050711}. The pair density wave is believed to exist in cuprate \cite{Berg_2009,PhysRevLett.93.187002,PhysRevLett.99.127003} and certain heavy Fermion superconductors \cite{gerber_switching_2014,PhysRevX.6.041059}. The gapless Higgs mode discussed here can readily be generalized to the pair density wave. Soto-Garrido \textit{et al.} studied the fluctuation associated with the amplitude of pair density wave, i.e. fluctuation of $\Psi_0$ in the order parameter $\Psi = \Psi_0 \cos(\mathbf{Q}\cdot\mathbf{r}) e^{i\phi}$. They found that the gapped Higgs mode is stable \cite{PhysRevB.95.214502}. They did not study explicitly the Higgs mode associated with the translation of the pair density wave, which is the main focus of the present work. The experimental observation of the gapless Higgs mode, at zero magnetic field, would also provide  a strong evidence for the pair density wave phase in the cuprates and heavy fermion systems.

{\it Conclusion ---} We have demonstrated the existence of a gapless Higgs mode in the FFLO state of an $s$-wave superconductor. The gapless Higgs mode originates from the translational symmetry breaking and is also the acoustic ``phonon" mode of the FFLO state. This gapless Higgs mode is stable because its decay into other modes is suppressed or forbidden kinematically. The linearly dispersive Higgs mode with zero gap shows up in thermodynamical quantities, i.e. it contributes a $T^3$ dependence term to the specific heat. The gapless Higgs mode can be probed by Raman spectroscopy and time-resolved terahertz spectroscopy techniques.   

{\it Acknowledgments ---}  Z.H. is grateful to Muneto Nitta, Weiqiang Chen, Shunji Tsuchiya and Ryosuke Yoshii for helpful discussions. This work was carried out under the auspices of the U.S. DOE NNSA under Contract No. 89233218CNA000001 through the LANL LDRD Program and the U.S. DOE Office of Basic Energy Sciences Program E3B5 (Z.H., S.-Z.L. \& J.-X.Z.). C.S.T. is supported by the Robert A. Welch foundation under grant no. E-1146. It was performed, in part, at the Center for Integrated Nanotechnologies, an Office of Science User Facility operated for the U.S. Department of Energy (DOE) Office of Science.  Computer resources for numerical calculations were supported by the Institutional Computing Program at LANL.

\begin{widetext}
\section{Action of Higgs mode}

We consider a two dimensional BCS superconductor under a Zeeman field described by the Hamiltonian
\begin{equation}
H=\sum_\sigma\int d^{2}r\psi_\sigma^\dagger\left(-\nabla^2/2m-\mu+E_z\sigma_z\right)\psi_\sigma-V\int d \tau d^{2}r \psi_\sigma^\dagger \psi_{\bar{\sigma}}^\dagger\psi_{\bar{\sigma}}\psi_\sigma,
\end{equation}
where $\psi_\sigma^\dagger(\psi_\sigma)$ is the electron creation (annihilation) operator with the chemical potential $\mu$, spin index $\sigma$ and pairing interaction strength $V$. The corresponding action in the imaginary time representation is
\begin{equation}
S=\int_0^\beta d\tau\sum_\sigma\int d^{2}r\left(\psi_\sigma^\dagger\partial_\tau\psi_\sigma+H(\tau,\mathbf{r})\right),
\end{equation}
where $\beta=1/k_BT$ with $T$ the temperature and $\mathbf{r}=(z,x)$. We then perform the Hubbard-Stratonivich transformation to introduce the superconducting energy gap $\Delta$ as
\begin{align}
\exp\left[ V\int d\tau d^2 r \psi_\sigma^\dagger \psi_{\bar{\sigma}}^\dagger\psi_{\bar{\sigma}}\psi_\sigma \right] = \\
\int D[\Delta,\Delta^*]\exp\left[-\int d\tau d^2 r\left( \frac{|\Delta|^2}{V}-\Delta^*\psi_{\bar{\sigma}}\psi_\sigma-\Delta\psi_\sigma^\dagger \psi_{\bar{\sigma}}^\dagger\right)\right].
\end{align}
By introducing the Nambu spinor operators $\Psi^\dagger=(\psi_\uparrow^\dagger, \psi_\downarrow)$, the action becomes \cite{Atland10}
\begin{equation}
S=-\int_0^\beta d\tau\int d^{2}r\Psi^\dagger G_0^{-1}
\Psi+\int_0^\beta d\tau\int d^{2}r\frac{|\Delta|^2}{V}
\end{equation}
with 
\begin{equation}
G_{0}^{-1}=-\left[\begin{array}{cc}
\partial_{\tau}+h+E_{z} & -\Delta\\
-\Delta & \partial_{\tau}-h+E_{z}
\end{array}\right]\;,
\end{equation}
and $h=-\nabla^2/2m -\mu$.
Integrating out the fermionic degree of freedom, we obtain
\begin{equation}
S=\int d\tau\int d^{2}r\left(\frac{1}{V}\left|\Delta\right|^{2}\right)-\text{tr}\ln G_{0}^{-1}.
\end{equation}
Now we include the amplitude fluctuation $s\left(\tau,\mathbf{r}\right)$ into
the action and obtain
\begin{equation}
S=\int d\tau\int d^{2}r\left(\frac{1}{V}\left|\Delta+s\right|^{2}\right)-\text{tr}\ln\left(G_{0}^{-1}+\Sigma\right)
\end{equation}
 with 
\[
\Sigma=s\sigma_{x}.
\]
Here $s$ is real and spatial dependent and much smaller than $\Delta$.
The action can be expanded to the second order of $s$ as
\begin{align} \label{totalaction}
S & =\int d\tau\int d^{2}r\left(\frac{1}{V}\left|\Delta\right|^{2}\right)-\text{tr}\ln G_{0}^{-1}\\
 & +\int d\tau\int d^{2}r\left(\frac{1}{V}2\Delta s\right)-\text{tr}\left(G_{0}\Sigma\right)\nonumber \\
 & +\int d\tau\int d^{2}r\left(\frac{1}{V}s^{2}\right)+\frac{1}{2}\text{tr}\left(G_{0}\Sigma G_{0}\Sigma\right),\nonumber 
\end{align}
where the terms related to first-order perturbation (second row in Eq.~\ref{totalaction}) are zero from condition of the saddle
point. We specifically take out the second-order perturbation 
\begin{equation}
S_{2}=\int d\tau\int d^{2}r\left(\frac{1}{V}s^{2}\right)+\frac{1}{2}\text{tr}\left(G_{0}\Sigma G_{0}\Sigma\right),\label{secondorderaction}
\end{equation}
which is the action describing the Higgs modes. 
In the frequency-momentum representation, we have
\begin{equation}
\int d\tau\int d^{2}r\left(\frac{1}{V}s^{2}\right)=\frac{1}{V}\Sigma_{l,\mathbf{q}}s\left(-\Omega_{l},-\mathbf{q}\right)s\left(\Omega_{l},\mathbf{q}\right)
\end{equation}
and
\begin{align}
tr\left(G_{0}\Sigma G_{0}\Sigma\right) & =\frac{T}{L^{2}}\Sigma_{m,l,\mathbf{k},q}tr\left(\left\langle \omega_{m},\mathbf{k}\right|G_{0}\left|\omega_{m},\mathbf{k}\right\rangle \left\langle \omega_{m},\mathbf{k}\right|\Sigma\left|\omega_{m}+\Omega_{l},\mathbf{k}+\mathbf{q}\right\rangle \right.\nonumber \\
 & \left.\left\langle \omega_{m}+\Omega_{l},\mathbf{k+q}\right|G_{0}\left|\omega_{m}+\Omega_{l},\mathbf{k+q}\right\rangle \left\langle \omega_{m}+\Omega_{l},\mathbf{k+q}\right|\Sigma\left|\omega_{m},\mathbf{k}\right\rangle \right)
\end{align}
with $\omega_{m}=\left(2m+1\right)\pi T$ and $\Omega_{l}=2l\pi T$ the fermionic and bosonic Matsubara frequency. To summarise, we
rewrite $S_{2}$ as
\begin{equation}
S_{2}=\Sigma_{l,\mathbf{q}}s\left(-\Omega_{l},\mathbf{-q}\right)Ms\left(\Omega_{l},\mathbf{q}\right)
\end{equation}
where the coupling matrix $M$ is given by
\begin{equation}
M=\frac{1}{V}+\frac{1}{2}\frac{T}{L^{2}}\Sigma_{m,\mathbf{k}}\text{tr}\left[G_{0}\left(\omega_{m},\mathbf{k}\right)\sigma_{x}G_{0}\left(\omega_{m}+\Omega_{l},\mathbf{k+q}\right)\sigma_{x}\right].\label{couplingmatrix}
\end{equation}

\section{Action of Higgs modes in uniform states}
By considering the uniform superconductivity, in the frequency-momentum representation, we have $G_{0}^{-1}=\left[\begin{array}{cc}
i\omega_{n}-h_{k}-E_z & \Delta\\
\Delta & i\omega_{n}+h_{k}-E_z
\end{array}\right]$ and 
\begin{equation}
G_{0}=\frac{i\omega_{n}-E_z+h_{k}\sigma_{z}-\Delta\sigma_{x}}{\left(i\omega_{n}-E_z\right)^{2}-h_{k}^{2}-\Delta^{2}}\label{G1}
\end{equation}
 with $h_{k}=\xi_{k}-\mu$. The saddle point is $\frac{\delta S}{\delta\Delta}=0$
which gives
\begin{equation}
\frac{2}{V}\Delta=\frac{T}{L^{2}}\Sigma_{k,n}\frac{-2\Delta}{(i\omega_{n}-E_z)^{2}-\lambda_k^2}=\frac{2\Delta}{L^{2}}\Sigma_{k}\frac{1-f\left(\lambda_{k}+E_z\right)-f\left(\lambda_{k}-E_z\right)}{2\lambda_{k}}\label{saddlecondition}
\end{equation}
where $\lambda_{k}=\sqrt{h_{k}^{2}+\Delta^{2}}$, $T$ is the
temperature, $L$ is the length of a side of a square sample and $f(E)$
indicates the Fermi-Dirac distribution. We consider zero temperature and the Zeeman field region where $|E_z|\le\Delta$. This gives the same self-consistent gap equation as the uniform superconductivity without the Zeeman field. Therefore $\Delta$ is independent of $E_z$ in this region.
We first calculate the trace
\begin{equation}
\mathrm{tr}\left[G_{0}\left(\omega_{m},k\right)\sigma_{x}G_{0}\left(\omega_{m}+\Omega_{l},k+q\right)\sigma_{x}\right]=\frac{\mathrm{tr}\left[\left(i\omega_{m}-E_z+h_{k}\sigma_{z}-\Delta\sigma_{x}\right)\sigma_{x}\left(i\omega_{m}-E_z+i\Omega_{l}+h_{k+q}\sigma_{z}-\Delta\sigma_{x}\right)\sigma_{x}\right]}{\left[\left(i\omega_{m}-E_z\right)^{2}-\lambda_{k}^{2}\right]\left[\left(i\omega_{m}-E_z+i\Omega_{l}\right)^{2}-\lambda_{k+q}^{2}\right]}.
\end{equation}
By using the trace of products of Pauli matrices given by 
\begin{equation}
\mathrm{tr}\left(\sigma_{i}\sigma_{j}\right)=2\delta_{ij},\ \mathrm{tr}\left(\sigma_{i}\sigma_{j}\sigma_{k}\right)=2i\varepsilon_{ijk},\ \mathrm{tr}\left(\sigma_{i}\sigma_{j}\sigma_{k}\sigma_{l}\right)=2\left(\delta_{ij}\delta_{kl}-\delta_{ik}\delta_{jl}+\delta_{il}\delta_{kj}\right),
\end{equation}
we obtain 
\begin{equation}
\mathrm{tr}\left[G_{0}\left(\omega_{m},k\right)\sigma_{x}G_{0}\left(\omega_{m}+\Omega_{l},k+q\right)\sigma_{x}\right]=2\frac{\Delta^{2}+(i\omega_{m}-E_z)\left(i\omega_{m}-E_z+i\Omega_{l}\right)-h_{k}h_{k+q}}{\left[\left(i\omega_{m}-E_z\right)^{2}-\lambda_{k}^{2}\right]\left[\left(i\omega_{m}-E_z+i\Omega_{l}\right)^{2}-\lambda_{k+q}^{2}\right]}.\label{trace1}
\end{equation}
By putting Eq. $\left(\ref{trace1}\right)$ into Eq. $\left(\ref{couplingmatrix}\right)$
and then sum over $\omega_{m}$, we get the coupling matrix
\begin{equation}
M=\frac{1}{V}+\frac{1}{2}\frac{1}{L^{2}}\prod\label{M}
\end{equation}
where
\begin{align}
\prod & =\sum_{k}\frac{\Delta^{2}+\lambda_{k}\left(\lambda_{k}+i\Omega_{l}\right)-h_{k}h_{k+q}}{\lambda_{k}\left[\left(\lambda_{k}+i\Omega_{l}\right)^{2}-\lambda_{k+q}^{2}\right]}f\left(\lambda_{k}+E_z\right)\label{trace2}\\
 & +\sum_{k}\frac{\Delta^{2}-\lambda_{k}\left(-\lambda_{k}+i\Omega_{l}\right)-h_{k}h_{k+q}}{-\lambda_{k}\left[\left(-\lambda_{k}+i\Omega_{l}\right)^{2}-\lambda_{k+q}^{2}\right]}f\left(-\lambda_{k}+E_z\right)\nonumber \\
 & +\sum_{k}\frac{\Delta^{2}+\lambda_{k+q}\left(\lambda_{k+q}-i\Omega_{l}\right)-h_{k}h_{k+q}}{\lambda_{k+q}\left[\left(\lambda_{k+q}-i\Omega_{l}\right)^{2}-\lambda_{k}^{2}\right]}f\left(\lambda_{k+q}+E_z\right)\nonumber \\
 & +\sum_{k}\frac{\Delta^{2}-\lambda_{k+q}\left(-\lambda_{k+q}-i\Omega_{l}\right)-h_{k}h_{k+q}}{-\lambda_{k+q}\left[\left(-\lambda_{k+q}-i\Omega_{l}\right)^{2}-\lambda_{k}^{2}\right]}f\left(-\lambda_{k+q}+E_z\right).\nonumber 
\end{align}
We also consider the zero temperature and $|E_z|\le\Delta$,
so $f\left(\lambda_{k}+E_z\right)=0$ and $f\left(-\lambda_{k}+E_z\right)=1$.
Eq. $\left(\ref{trace2}\right)$ is rewritten as 
\begin{align}
 & \mathrm{tr}\left[G_{0}\left(\omega_{m},k\right)\sigma_{x}G_{0}\left(\omega_{m}+\Omega_{l},k+q\right)\sigma_{x}\right]\nonumber \\
 & =\sum_{k}-\left\{ \frac{\Delta^{2}-\lambda_{k}\left(-\lambda_{k}+i\Omega_{l}\right)-h_{k}h_{k+q}}{\lambda_{k}\left[\left(-\lambda_{k}+i\Omega_{l}\right)^{2}-\lambda_{k+q}^{2}\right]}+\frac{\Delta^{2}+\lambda_{k+q}\left(\lambda_{k+q}+i\Omega_{l}\right)-h_{k}h_{k+q}}{\lambda_{k+q}\left[\left(\lambda_{k+q}+i\Omega_{l}\right)^{2}-\lambda_{k}^{2}\right]}\right\} 
\end{align}
which is the same as the case $E_z=0$. Since $\Delta$ is also the same as the case $E_z=0$, we have $M_q$ the same as the case $E_z=0$ in the region $|E_z|\le\Delta$. Therefore the Higgs mode dispersion is the same as that at $E_z=0$, which is $\Omega_H^2=4\Delta^2+\frac{1}{3}v_F^2 |\mathbf{q}|^2$ \cite{Littlewood82}.

\section{Action of Higgs modes in FFLO states}

To find the FFLO solution, we switch to a tight binding model which is more convenient for numerical calculations. The tight-binding Hamiltonain with the Zeeman term is given by
\begin{equation}
H=\sum_{\mathbf{r}}-\mu\psi_{\mathbf{r\sigma}}^{\dagger}\psi_{\mathbf{r\sigma}}+E_{z}\psi_{\mathbf{r}\uparrow}^{\dagger}\psi_{\mathbf{r}\uparrow}-E_{z}\psi_{\mathbf{r\downarrow}}^{\dagger}\psi_{\mathbf{r\downarrow}}-t\psi_{\mathbf{r\sigma}}^{\dagger}\psi_{\mathbf{r+\delta,\sigma}}+\Delta(\mathbf{r})\psi_{\mathbf{r}\uparrow}^{\dagger}\psi_{\mathbf{r}\downarrow}^{\dagger}+\Delta(\mathbf{r})^{*}\psi_{\mathbf{r}\downarrow}\psi_{\mathbf{r}\uparrow},
\end{equation}
with $\mathbf{r}=\left(z,x\right)$. We consider the FFLO solution which breaks (preserves) the translational symmetry in the $x$ ($z$) direction. Then in the $x$ direction, there are $l_x$
supercells each with $N_{x}$ meshes and the total length is $L_{x}=l_{x}N_{x}$. In the momentum space, the Hamiltonian is given by 
\paragraph{
\[
\mathscr{H}=\sum_{k_{z},k_{x}}\Psi^{\dagger}\tilde{H}\Psi,
\]}
where $\Psi^{\dagger}=\left[\Psi_{\uparrow k_{z},k_{x}x_{1}}^{\dagger},\Psi_{\downarrow-k_{z},-k_{x},x_{1}},\Psi_{\uparrow k_{z},k_{x},x_{2}}^{\dagger},\Psi_{\downarrow-k_{z},-k_{x},x_{2}}...,\Psi_{\uparrow k_{z},k_{x},x_{N}}^{\dagger},\Psi_{\downarrow-k_{z},-k_{x},x_{N}}\right]$
and
\[
\tilde{H}=\left[\begin{array}{ccccccc}
h+E_z & \Delta & -t & 0 & \cdots & -te^{-ik_{x}N_x} & 0\\
\Delta^{*} & -h+E_z & 0 & t & \cdots & 0 & te^{-ik_{x}N_x}\\
-t & 0 & h+E_z & \Delta & \cdots & 0 & 0\\
0 & t & \Delta^{*} & -h+E_z & \cdots & 0 & 0\\
\vdots & \vdots & \vdots & \vdots & \ddots & \vdots & \vdots\\
-te^{ik_{x}N_x} & 0 & 0 & 0 & \cdots & h+E_z & \Delta\\
0 & te^{ik_{x}N_x} & 0 & 0 & \cdots & \Delta^{*} & -h+E_z
\end{array}\right].
\]
with $h=-2t\cos k_z-\mu$, $k_{x}=\frac{2\pi}{L_{x}}l_{x}$ and $k_{z}=\frac{2\pi}{L_{z}}m_{z}$. 
By using the canonical transformation
\begin{align}
\Psi_{\mathbf{k}}\left(x_{i}\right) & =\sum_{n}u_{\uparrow \mathbf{k}}^{n_{1}}\left(x_{i}\right)\gamma_{n1}-v_{\uparrow \mathbf{k}}^{*n_{2}}\left(x_{i}\right)\gamma_{n2}^{\dagger},\\
\Psi_{\uparrow \mathbf{k}}^{\dagger}\left(x_{i}\right) & =\sum_{n}u_{\uparrow\mathbf{k}}^{*n_{1}}\left(x_{i}\right)\gamma_{n1}^{\dagger}-v_{\uparrow \mathbf{k}}^{n_{2}}\left(x_{i}\right)\gamma_{n2},\\
\Psi_{\downarrow-\mathbf{k}}\left(x_{i}\right) & =\sum_{n}u_{\downarrow-\mathbf{k}}^{n_{2}}\left(x_{i}\right)\gamma_{n2}+v_{\downarrow-\mathbf{k}}^{*n_{1}}\left(x_{i}\right)\gamma_{n1}^{\dagger},\\
\Psi_{\downarrow-\mathbf{k}}^{\dagger}\left(x_{i}\right) & =\sum_{n}u_{\downarrow-\mathbf{k}}^{*n_{2}}\left(x_{i}\right)\gamma_{n2}^{\dagger}+v_{\downarrow-\mathbf{k}}^{n_{1}}\left(x_{i}\right)\gamma_{n1},
\end{align}
the gap function is given by
\begin{equation}
\Delta =\frac{1}{L_zl_x}\sum_{\mathbf{k}}V\left\langle \Psi_{\uparrow \mathbf{k}}\left(x_{i}\right)\Psi_{\downarrow-\mathbf{k}}\left(x_{i}\right)\right\rangle =\frac{V}{L_zl_x}\sum_{\mathbf{k},n}u_{\uparrow \mathbf{k}}^{n}\left(x_{i}\right)v_{\downarrow-\mathbf{k}}^{*n}\left(x_{i}\right)f\left(-E_{n}\right).
\end{equation}
For proper $E_z$, with self-consistent calculation we can obtain an FFLO state with one period in the supercell. The FFLO state has lower energy than the uniform superconducting and normal state in the intermediate Zeeman field region \cite{Ting06}. 

Now we write the Gorkov's Green function in the form \cite{Zhu_2016}
\begin{align}
G_{k}=\left[\begin{array}{cc}
G_{11} & G_{12}\\
G_{21} & G_{22}
\end{array}\right],
\end{align}
where
\begin{align}
G_{11} & =\int_{0}^{\beta}d\tau\ e^{i\omega_{n}\tau}\left[-\left\langle \Psi_{\uparrow \mathbf{k}}\left(x_{1}\right)\Psi_{\uparrow \mathbf{k}}^{\dagger}\left(x_{2}\right)\right\rangle \right]\nonumber \\
 & =\sum_{n}\frac{u_{\uparrow \mathbf{k}}^{n}\left(x_{1}\right)u_{\uparrow \mathbf{k}}^{*n}\left(x_{2}\right)}{i\omega_{n}-E_{\mathbf{k}}^{n}},
\end{align}
and
\begin{align}
G_{12} & =\int_{0}^{\beta}d\tau\ e^{i\omega_{n}\tau}\left[-\left\langle \Psi_{\uparrow \mathbf{k}}\left(x_1\right)\Psi_{\downarrow-\mathbf{k}}\left(x_2\right)\right\rangle \right]\\
&=\sum_{n}\frac{u_{\uparrow \mathbf{k}}^{n}\left(x_{1}\right)v_{\downarrow-\mathbf{k}}^{*n}\left(x_{2}\right)}{i\omega_{n}-E_{\mathbf{k}}^{n}},
\end{align}
and
\begin{align}
 G_{21} & =\int_{0}^{\beta}d\tau\ e^{i\omega_{n}\tau}\left[-\left\langle \Psi_{\downarrow-\mathbf{k}}^{\dagger}\left(x_{1}\right)\Psi_{\uparrow \mathbf{k}}^{\dagger}\left(x_{2}\right)\right\rangle \right]\\
 & =\sum_{n}\frac{v_{\downarrow-\mathbf{k}}^{n}\left(x_{1}\right)u_{\uparrow \mathbf{k}}^{n*}\left(x_{2}\right)}{i\omega_{n}-E_{\mathbf{k}}^{n}},
\end{align}
and
\begin{align}
G_{22} & =\int_{0}^{\beta}d\tau\ e^{i\omega_{n}\tau}\left[-\left\langle \Psi_{\downarrow-\mathbf{k}}^{\dagger}\left(x_{1}\right)\Psi_{\downarrow-\mathbf{k}}\left(x_{2}\right)\right\rangle \right]\nonumber \\
 & =\sum_{n}\frac{v_{-\mathbf{k}}^{n}\left(x_{1}\right)v_{-\mathbf{k}}^{n*}\left(x_{2}\right)}{i\omega_{m}-E_{\mathbf{k}}^{n}}.
\end{align}
We thus further obtain
\begin{align}
 & \mathrm{tr}\left(G_{0}\Sigma G_{0}\Sigma\right)\nonumber \\
 & =\frac{T}{L_{z}l_{x}}\Sigma_{m,l,k_{x},q_{x},k_{z},q_{z,}x_{1},x_{2}}tr\left(\left\langle \omega_{m},k_{x},k_{z},x_{1}\right|G_{0}\left|\omega_{m},k_{x},k_{z},x_{2}\right\rangle \left\langle \omega_{m},k_{x},k_{z},x_{2}\right|\Sigma\left|\omega_{m}+\Omega_{l},k_{x}+q_{x},k_{z}+q_{z},x_{2}\right\rangle \right.\nonumber \\
 & \left.\left\langle \omega_{m}+\Omega_{l},k_{x}+q_{x},k_{z}+q_{z},x_{2}\right|G_{0}\left|\omega_{m}+\Omega_{l},k_{x}+q_{x},k_{z}+q_{z},x_{1}\right\rangle \left\langle \omega_{m}+\Omega_{l},k_{x}+q_{x},k_{z}+q_{z},x_{1}\right|\Sigma\left|\omega_{m},k_{x},k_{z},x_{1}\right\rangle \right)
\end{align}
where
\begin{equation}
\left\langle \omega_{m},k_{x},k_{z},x_{1}\right|G_{0}\left|\omega_{m},k_{x},k_{z},x_{2}\right\rangle =\sum_{n}\frac{1}{i\omega_{m}-E_{k_{x},k_{z}}^{n}}\left[\begin{array}{cc}
u_{k_{x},k_{z}}^{n}\left(x_{1}\right)u_{k_{x},k_{z}}^{n*}\left(x_{2}\right) & u_{k_{x},k_{z}}^{n}\left(x_{1}\right)v_{-k_{x},-k_{z}}^{n*}\left(x_{2}\right)\\
v_{-k_{x},-k_{z}}^{n}\left(x_{1}\right)u_{k_{x},k_{z}}^{n*}\left(x_{2}\right) & v_{-k_{x},-k_{z}}^{n}\left(x_{1}\right)v_{-k_{x},-k_{z}}^{n*}\left(x_{2}\right)
\end{array}\right],
\end{equation}
and
\begin{equation}
\left\langle \omega_{m},k_{x},k_{z},x_{2}\right|\Sigma\left|\omega_{m}+\Omega_{l},k_{x}+q_{x},k_{z}+q_{z},x_{2}\right\rangle =\left[\begin{array}{cc}
0 & s\left(-\Omega_{l},-q_{x},-q_{z},x_{2}\right)\\
s\left(-\Omega_{l},-q_{x},-q_{z},x_{2}\right) & 0
\end{array}\right],
\end{equation}
and
\begin{align}
 & \left\langle \omega_{m}+\Omega_{l},k_{x}+q_{x},k_{z}+q_{z},x_{2}\right|G_{0}\left|\omega_{m}+\Omega_{l},k_{x}+q_{x},k_{z}+q_{z},x_{1}\right\rangle \nonumber \\
 & =\sum_{d}\frac{1}{i\omega_{m}+i\Omega_{l}-E_{k_{x}+q_{x},k_{z}+q_{z}}^{d}}\left[\begin{array}{cc}
u_{k_{x}+q_{x},k_{z}+q_{z}}^{d}\left(x_{2}\right)u_{k_{x}+q_{x},k_{z}+q_{z}}^{d*}\left(x_{1}\right) & u_{k_{x}+q_{x},k_{z}+q_{z}}^{d}\left(x_{2}\right)v_{-\left(k_{x}+q_{x}\right),-\left(k_{z}+q_{z}\right)}^{d*}\left(x_{1}\right)\\
v_{-\left(k_{x}+q_{x}\right),-\left(k_{z}+q_{z}\right)}^{d}\left(x_{2}\right)u_{k_{x}+q_{x},k_{z}+q_{z}}^{d*}\left(x_{1}\right) & v_{-\left(k_{x}+q_{x}\right),-\left(k_{z}+q_{z}\right)}^{d}\left(x_{2}\right)v_{-\left(k_{x}+q_{x}\right),-\left(k_{z}+q_{z}\right)}^{d*}\left(x_{1}\right)
\end{array}\right]
\end{align}
and
\begin{equation}
\left\langle \omega_{m}+\Omega_{l},k_{x}+q_{x},k_{z}+q_{z},x_{1}\right|\Sigma\left|\omega_{m},k_{x},k_{z},x_{1}\right\rangle =\left[\begin{array}{cc}
0 & s\left(\Omega_{l},q_{x},q_{z},x_{1}\right)\\
s\left(\Omega_{l},q_{x},q_{z},x_{1}\right) & 0
\end{array}\right].
\end{equation}

By summarizing over $\omega_{m}$, we obtain
\begin{align}
\mathrm{tr}\left(G_{0}\Sigma G_{0}\Sigma\right) & =\frac{T}{L_{z}l_{x}}\sum_{m,l,k_{x},k_{y},q_{x},q_{z},x_{1},x_{2}}\sum_{n}\frac{s\left(-\Omega_{l},-\mathbf{q},x_{2}\right)}{i\omega_{m}-E_{k_{x},k_{z}}^{n}}\sum_{d}\frac{s\left(\Omega_{l},\mathbf{q},x_{1}\right)}{i\omega_{m}+i\Omega_{l}-E_{\mathbf{k}+\mathbf{q}}^{d}}\nonumber \\
 & \mathrm{tr}\left\{ \left[\begin{array}{cc}
u_\mathbf{k}^{n}\left(x_{1}\right)v_\mathbf{-k}^{n*}\left(x_{2}\right) & u_{\mathbf{k}}^{n}\left(x_{1}\right)u_{\mathbf{k}}^{n*}\left(x_{2}\right)\\
v_{-\mathbf{k}}^{n}\left(x_{1}\right)v_{-\mathbf{k}}^{n*}\left(x_{2}\right) & v_{-\mathbf{k}}^{n}\left(x_{1}\right)u_{\mathbf{k}}^{n*}\left(x_{2}\right)
\end{array}\right]\left[\begin{array}{cc}
u_{\mathbf{k+q}}^{d}\left(x_{2}\right)v_{-\left(\mathbf{k+q}\right)}^{d*}\left(x_{1}\right) & u_{\mathbf{k+q}}^{d}\left(x_{2}\right)u_{\mathbf{k+q}}^{d*}\left(x_{1}\right)\\
v_{-\left(\mathbf{k+q}\right)}^{d}\left(x_{2}\right)v_{-\left(\mathbf{k+q}\right)}^{d*}\left(x_{1}\right) & v_{-\left(\mathbf{k+q}\right)}^{d}\left(x_{2}\right)u_{\mathbf{k+q}}^{d*}\left(x_{1}\right)
\end{array}\right]\right\} \nonumber \\
\nonumber \\
 & =\frac{1}{L_{z}l_{x}}\sum_{l,\mathbf{q}}\sum_{x_{1},x_{2}}s\left(-\Omega_{l},-\mathbf{q},x_{2}\right)s\left(\Omega_{l},\mathbf{q},x_{1}\right)\sum_{\mathbf{k},n,d}\frac{D}{E_{\mathbf{k}}^{n}+i\Omega_{l}-E_{\mathbf{k+q}}^{d}}\left[f\left(E_{\mathbf{k}}^{n}\right)-f\left(E_{\mathbf{k+q}}^{d}\right)\right],
\end{align}
with 
\begin{align}
D & =u_{\mathbf{k}}^{n}\left(x_{1}\right)v_{-\mathbf{k}}^{n*}\left(x_{2}\right)u_{\mathbf{k+q}}^{d}\left(x_{2}\right)v_{-\left(\mathbf{k+q}\right)}^{d*}\left(x_{1}\right)+u_{\mathbf{k}}^{n}\left(x_{1}\right)u_{\mathbf{k}}^{n*}\left(x_{2}\right)v_{-\left(\mathbf{k+q}\right)}^{d}\left(x_{2}\right)v_{-\left(\mathbf{k+q}\right)}^{d*}\left(x_{1}\right)\nonumber \\
 & +v_{-\mathbf{k}}^{n}\left(x_{1}\right)v_{-\mathbf{k}}^{n*}\left(x_{2}\right)u_{\mathbf{k+q}}^{d}\left(x_{2}\right)u_{\mathbf{k+q}}^{d*}\left(x_{1}\right)+v_{-\boldsymbol{k}}^{n}\left(x_{1}\right)u_{\mathbf{k}}^{n*}\left(x_{2}\right)v_{-\left(\mathbf{k+q}\right)}^{d}\left(x_{2}\right)u_{\mathbf{k+q}}^{d*}\left(x_{1}\right).
\end{align}
After performing the Wick rotation $i\Omega_l\rightarrow \Omega+i\eta$ with $\eta$ being an infinitesimal positive number, we write the action for the amplitude fluctuations as
\begin{equation}
S_{2}\left(\Omega,\mathbf{q}\right)=\Gamma^{\dagger}\left(-\Omega,-\mathbf{q}\right)M_{\Omega,\mathbf{q}}\Gamma\left(\Omega,\mathbf{q}\right)
\end{equation}
with $\Gamma\left(\Omega,\mathbf{q}\right)=\left[s\left(\Omega,\mathbf{q},x_{1}\right),s\left(\Omega,\mathbf{q},x_{2}\right),\cdots,s\left(\Omega,\mathbf{q},x_{N_{x}}\right)\right]^{\mathrm{\dagger}}$ 
and
\begin{align}
M_{\Omega,\mathbf{q}} =\frac{1}{V}+\frac{1}{2L_{z}l_{x}}\sum_{\mathbf{k},n,d}\frac{D}{E_{\mathbf{k}}^{n}+\Omega+i\eta-E_{\mathbf{k+q}}^{d}}\left[f\left(E_{\mathbf{k}}^{n}\right)-f\left(E_{\mathbf{k+q}}^{d}\right)\right].
\end{align}

\end{widetext}

\bibliography{references}

\end{document}